%% file: main.tex
\documentclass[sigconf]{acmart}

\usepackage[utf8]{inputenc}
\usepackage{color}
\usepackage{kotex}
\usepackage[T1]{fontenc}
\usepackage{url}
\usepackage{adjustbox}

\usepackage{booktabs} 
\usepackage{multirow}
\usepackage{color}
\usepackage[]{todonotes}
\usepackage{algorithm}
\usepackage[noend]{algpseudocode}
\usepackage{hyperref}
\usepackage[hyphenbreaks]{breakurl}
\usepackage{caption}
\usepackage{subcaption}
\usepackage{makecell}

\setcopyright{none}

\acmConference{Advances in Financial Technology (AFT19)}
\acmYear{October 2019}
\copyrightyear{\textit{Zurich, Switzerland}}

\acmDOI{Forthcoming}
\acmISBN{}
\acmPrice{}

\def\bitcoin{Bitcoin}

\def\nEmails{4,360,575}

\def\nBasketsStepTwo{1,810}

\def\nBasketsStepFour{96}
\def\nCampaignsFirst{19}
\def\nCampaignsSecond{35}
\def\nEmailsStepFour{4,342,380}
\def\nAddresses{12,533}
\def\nExplicitAddressesActive{245}

\def\nAddressesActive{485}
\def\nClusters{52}
\def\nClustersSextortion{48}
\def\nAddressesNotInClustersSextortion{33}
\def\nAddressesNotInClusters{49}
\def\nAddressesInClusters{196}

\def\nClustersWithMultpleBaskets{50}

\def\nBasketsGroupedInCampaigns{80}
\def\nEmailsConsidered{4,340,736}
\def\StartCollectionPeriod{2018-10-02}
\def\EndCollectionPeriod{2019-02-13}

\def\nTxsSecondFilter{2,252}

\def\revenuesSecondFilter{1,352,266}
\def\revenuesThirdFilter{1,300,620}
\def\revenuesOctober{336,979}
\def\revenuesNovember{286,462}
\def\revenuesJanuary{250,189}
\def\averageMonthlyRevenue{122,933} 

\begin{document}

\title{Spams meet Cryptocurrencies: Sextortion in the Bitcoin Ecosystem}

\author{Masarah Paquet-Clouston}
\affiliation{\institution{GoSecure}}
\email{mcpc@gosecure.net}

\author{Matteo Romiti}
\affiliation{\institution{Austrian Institute of Technology}}
\email{Matteo.Romiti@ait.ac.at}

\author{Bernhard Haslhofer}
\affiliation{\institution{Austrian Institute of Technology}}
\email{Bernhard.Haslhofer@ait.ac.at}

\author{Thomas Charvat}
\affiliation{\institution{Excello}}
\email{tc@excello.cz}


\begin{abstract}
In the past year, a new spamming scheme has emerged: sexual extortion messages requiring payments in the cryptocurrency \bitcoin{}, also known as \emph{sextortion}. This scheme represents a first integration of the use of cryptocurrencies by members of the spamming industry. Using a dataset of \nEmailsConsidered~ sextortion spams, this research aims at understanding such new amalgamation by uncovering spammers' operations. To do so, a simple, yet effective method for projecting \bitcoin{} addresses mentioned in sextortion spams onto transaction graph abstractions is computed over the entire \bitcoin{} blockchain. This allows us to track and investigate monetary flows between involved actors and gain insights into the financial structure of sextortion campaigns. We find that sextortion spammers are somewhat sophisticated, following pricing strategies and benefiting from cost reductions as their operations cut the upper-tail of the spamming supply chain. We discover that one single entity is likely controlling the financial backbone of the majority of the sextortion campaigns and that the 11-month operation studied yielded a lower-bound revenue between \$\revenuesThirdFilter{} and \$\revenuesSecondFilter{}. We conclude that sextortion spamming is a lucrative business and spammers will likely continue to send bulk emails that try to extort money through cryptocurrencies.

\end{abstract}

\maketitle


\input{sections/introduction}
\input{sections/relatedwork}
\input{sections/dataset}
\input{sections/analysis}
\input{sections/discussion}
\input{sections/conclusions}

\section*{Acknowledgments}

Work on this topic is supported inter alia by the European Union's Horizon 2020 research
and innovation programme under grant agreement No. 740558 (TITANIUM) and the Austrian FFG’s KIRAS programme under project VIRTCRIME (No. 860672).

\bibliographystyle{acm}

\bibliography{Bibliography} 

\end{document}

%% file: sections/introduction.tex

\section{Introduction}\label{sec:introduction}

In the past year, a new spamming scheme has emerged: sexual extortion messages requiring payments in the cryptocurrency Bitcoin, known as \textit{sextortion} spam. The threat is about the disclosure of compromising images and/or videos to the recipient’s contacts if an amount of money is not sent to a specific \bitcoin{} address. This sextortion spam scheme mixes a well-known existing online problem, spams,~\cite{rao2012,stonegross2011,kanich2011, kanich2008,stringhini2014}, while leveraging \textit{cryptocurrencies} as a new channel for illicit financial transactions. It represents a first integration of the use of cryptocurrencies by members of the spamming industry. Such spamming scheme is much simpler than conventional spamming that requires, for example, links to affiliate marketing~\cite{rao2012} for the sale of goods such as pharmaceutical products~\cite{kanich2008,kanich2011} or counterfeit goods~\cite{stringhini2014}. Indeed, sextortion spams cut the upper-tail of the spamming supply chain: there are no product shipments or hosting of illicit websites involved.

Since 2018, sextortion spams have been distributed in a dozen languages, most likely with the use of the Necurs botnet~\cite{schultz2018, ibmx-forceexchange2018}. Shultz~\cite{schultz2018} already conducted a primary analysis of the sextortion spammers potential revenues by inspecting 58,611 \bitcoin{} addresses found in two sextortion-related spam campaigns lasting 60 days. He concluded that a total of 83 addresses received approximately \$146,280. However, the author only summed incoming payments without applying more advanced methods for tracing monetary flows in the \bitcoin{} transaction graph, which have already been applied for studying other forms of cryptocurrency-related cybercrime, such as ransomware~\cite{paquet2018,huang2018}. Shultz's~\cite{schultz2018} findings provide a first overview of a much bigger puzzle, which has yet to be uncovered and understood. Moreover, the simplicity behind sextortion spams --- or any spams taking advantage of cryptocurrencies --- may influence many to leverage this strategy, developing new threats and/or offering different products or services to convince spam recipients to send money to a cryptocurrency address. Uncovering the potential returns of this scheme, as well as the level of sophistication behind these campaigns, may indicate the extent to which this illicit scheme will be prevalent in the near future.

This research builds on unique and innovative methods to understand this new spamming scheme, providing insights into how sextortion spammers operate and the extent to which they are successful. To do so, we dive into a dataset of \nEmailsConsidered~ emails related to sextortion, identify different threat campaigns and evaluate spammers' pricing strategy. Then, we extract \bitcoin{} addresses and investigate transactions related to sextortion in the \bitcoin{} ecosystem. Using the open-source \textit{GraphSense Cryptocurrency Analytics Platform}\footnote{https://graphsense.info/}, we project them onto transaction graphs by applying the well-known multiple-input clustering heuristic~\cite{Meiklejohn:2013a,Reid:2013a}. We estimate spammers' potential revenue through different techniques and analyze patterns in the monetary flows. Summarizing, our contributions are as follows:

\begin{enumerate}

    \item We develop a simple and efficient \textit{email bucketing heuristic} to syntactically group \nEmailsConsidered{} emails into \nBasketsStepFour{} sextortion-related buckets, from which we manually derive \nCampaignsSecond{} specific campaigns, allowing us to overcome spammers' obfuscation techniques of changing words and sentences.
    
    \item We find that sextortion spammers modify the amounts asked in spams, based on the language in which the spam is sent and the campaign in which it is included. 
    
    
    \item From analyzing the \nExplicitAddressesActive~ sextortion addresses active in the \bitcoin{} blockchain, which we make available for research reproducibility\footnote{https://github.com/MatteoRomiti/Sextortion\_Spam\_Bitcoin}, we find that one single entity is likely controlling the financial backbone of the majority of the sextortion campaigns studied.
    
    \item Filtering out unrelated transactions, we conclude that this sextortion scheme generated a lower-bound revenue between \$\revenuesThirdFilter{} and \$\revenuesSecondFilter{} for an 11-month period.
    
	\item We find that the holding period of sextortion payments is, on average, 5.5 days and we track, to the best of our capacity, monetary flows from spammers to known entities. 
	
\end{enumerate}

Our results show that sextortion spammers are somewhat sophisticated, developing pricing strategies before sending spams, yet they are not tracking whether the developed strategies worked, as \bitcoin{} addresses are reused among spams. We also find that the lower-bound revenues of sextortion spams are similar in magnitude compared to traditional spams~\cite{kanich2008, kanich2011}. However, by cutting the upper-tail of the spamming supply chain, sextortion spammers make near \$\averageMonthlyRevenue{} per month while facing little costs. These results suggest that sextortion spams, or any spamming scheme using cryptocurrencies to attempt at influencing spam recipients to blindly send money, are not likely to disappear, due to the lucrative returns. The anti-spam community should thus focus on actively flagging these emails, preventing them to reach users' inbox. We make our research reproducible by publishing the \nAddresses{} \bitcoin{} addresses found in the \nEmailsConsidered{} sextortion spams\textsuperscript{2}. 

In the following paper, we start by introducing the necessary background on sextortion spams, summarizing what is currently known about their characteristics and the profitability of the spamming industry, and providing an overview of known methods for estimating illicit revenues using \bitcoin{}. Then, in Section~\ref{sec:dataset}, we describe the simple and efficient \textit{email bucketing heuristic} we developed to group emails in buckets and subsequently campaigns, as well as the computations done on known \bitcoin{} addresses to expand the dataset and infer additional information from the \bitcoin{} ecosystem. We present results about the spammers' pricing strategy, the reuse of \bitcoin{} addresses, the evolution of spamming revenues as well as an analysis of the money flows in the subsequent sections. We finish with a discussion and a conclusion summarizing the main results of this study.

%% file: sections/relatedwork.tex
\section{Background and Related Work}\label{sec:relatedwork}

Since sextortion is a specific spamming scheme, we first summarize what is currently known about spam and the industry behind it in general, before we present known facts about spam sextortion. Then, we focus on profitability of spam and summarize known methods for measuring illicit cryptocurrency revenues.

\subsection{Spam and the Industry behind it}\label{sec:relatedwork_industry}

Spam refers to the sending of in-mass ``messages from economic agents who do not have previous relationship with the customer and who do not offer opt-out provisions''~\cite[p.96]{rao2012} or, put more simply, ``unsolicited bulk emails''~\cite{androutsopoulos2000}. Spam was first attributed to unwanted emails, but now refers to other kinds of undesirable messages, such as unsolicited text messages~\cite{chen2015}, calls~\cite{tu2016}, social media posts~\cite{chen2015,sedhai2015} and fake reviews~\cite{heydari2015}.

The spamming industry, referring to all economic agents involved in the spamming business, has been thriving since the 1980’s with the creation of the Simple Mail Transfer Protocol (SMTP) sender push technology~\cite{rao2012}. Since its inception, this industry has been closely related to affiliate marketing~\cite{rao2012}, advertising various products and services, such as pharmaceutical products~\cite{kanich2011,kanich2008}, counterfeit goods, online dating and search engine optimization services~\cite{stringhini2014}.

Stringhini et al.~\cite{stringhini2014} suggested that the lower-tail of the spamming supply chain is segmented, with each participant specializing in one activity. They also point out that traditional spammers buy emails lists, rent botnets and use them to send spam. Target email lists are bought from email harvesters, which are individuals or groups of individuals crawling the web and compiling large lists of emails for sale. Then, botnets would be rented by spammers to send their unsolicited emails~\cite{john2009,kanich2011,stonegross2011}. Botnets enable spam to be sent in bulk from a vast pool of infected hosts with geographic diversity, especially if the infected host is located in a Dynamic Host Protocol (DHCP). In such case, the IP address of the infected host may be constantly changing, thus hardening the flagging process by the anti-spam community~\cite{rao2012}. 

While Stringhini et al.~\cite{stringhini2014} suggested a segmented industry at the lower-tail, Kanich et al.~\cite{kanich2008} investigated the Storm spam botnet and concluded that the spamming industry would be more vertically integrated, with the harvesters, the botnet owners and the spammers coming from the same organization. 

Then, a study conducted by Levenchenko et al.~\cite{levchenko2011} showed that, at the upper tail of the traditional spamming supply chain, several operations need to be performed. The authors discussed, more precisely, spam campaigns related to pharmaceutical and software products and explained that sending spam is only a small part of the whole operation. Once in a virtual inbox, spammers depend on the propensity of email recipients to click on the URL displayed in the message and for the URL to remain available. Although this process seems simple at first, Levenchenko et al. ~\cite{levchenko2011} explain that spammers need to orchestrate and plan an elaborated operation to avoid URL and domain blacklisting, as well as website take downs by Internet Service Providers (ISPs) and registrars. Such an operation is quite sophisticated and requires URL redirection strategies, dealing with bullet-proof hosters and even sometimes using fast flux domain name system (DNS). Moreover, spammers must maintain credible websites, provide reliable payment processes and find banks that will accept to receive payments, regardless of the illicit activity being conducted. Lastly, they may have to acquire the product and deliver it to the customer and/or develop the service. Since the supply chain for the spamming of counterfeit products and software services is quite complex and requires sophisticated operations, spammers can also resort to enterprises with lenient values to conduct business with, sending spams on their behalf \cite{rao2012}.

\subsection{Spam Sextortion}\label{sec:spam_sextortion}

With the rise of \bitcoin{} as a preferred currency for specific illicit activities, such as ransomware~\cite{paquet2018,huang2018}, some spammers have returned to spamming through emails while exploiting a new strategy: sexual extortion that requires payments in \bitcoin{}. The scheme, known as \textit{spam sextortion}, is simple: it aims at extorting money from spam recipients through the threat that compromising photos or videos will be sent to the recipients' contacts if the amount asked in \bitcoin{} is not paid. Such spamming scheme represents a new approach that leverages \bitcoin{} and is cost-saving: it cuts the spamming supply chain to its bare minimal as there is no need to develop sophisticated techniques to monetize the scheme.

Indeed, the upper-tail of the supply chain is butchered when considering sextortion spam campaigns: the spam is built on a false threat that if believed by the spam recipient, will generate an amount of money to the spammers. There are no URL redirections and bulletproof hosting to deal with, credible websites to maintain, or business partners to find. The cost of running a sextortion spam campaign is thus largely reduced compared to traditional spams, which may explain why, in the past year, such spams have been prevalent. To this date, sextortion spams have been discussed in mid-2018 in a short article from IBM X-Force ~\cite{ibmx-forceexchange2018} as well as a security blog written by Schultz (2018) ~\cite{schultz2018}. We review both articles below. 

The first short article~\cite{ibmx-forceexchange2018} presents examples of spam sextortion and hypothesizes that these spams have been distributed since September 11, 2018, by the Necurs botnet, a well-known spamming botnet~\cite{Kessem}. According to the blog, the amounts asked varied between \$250 and \$550 and the threat was found to be distributed in seven languages: English, German, French, Italian, Japanese, Korean and Arabic. 
The second article, written by Schultz~\cite{schultz2018} in 2018, investigated two sextortion campaigns (a total of 233,336 emails) that were received by SpamCop, a reporting spam service~\footnote{https://www.spamcop.net/}, between August 30 and October 26, 2018. The author found additional languages: Czech, Spanish, Norwegian, Swedish and Finnish, illustrating that the sextortion spamming scheme is inevitably global. Schultz also found that \bitcoin{} addresses displayed in spams are reused throughout sextortion messages: the spammers do not even bother to generate a unique address per message. This is further supported in the previous article~\cite{ibmx-forceexchange2018} which stated that one \bitcoin{} address was reused in 3 million messages. Schultz \cite{schultz2018}, moreover, found overlaps of \bitcoin{} addresses in sextortion spam emails and a spam where a Russian girl promised to send an explicit video in exchange for \$40 in \bitcoin{}s.

These findings suggest that there might be a common financial backbone behind the spam campaigns. However, a more detailed picture on the structure of that backbone and revenues made by sextortion spammers are still missing. But, since spammers clearly tend to reuse addresses, we can infer and show later in this paper that such a picture can be drawn using state-of-the-art methods for tracing and tracking monetary flows in \bitcoin{}.

\subsection{The Potential Profitability of Spams}\label{sec:relatedwork_profitability}

Traditional spamming is a lucrative business: in 2008, Kanich et al.~\cite{kanich2008} infiltrated a botnet for 26 days and found that 350 million pharmaceutical spams resulted in 28 sales with a total revenue of \$2,731.88, which corresponds to a conversion rate of 0.00001\%. The authors claim that, when extrapolating their findings to additional bots, the revenue could go up to \$9,500 per day and thus \$3.5 million per year. Later, the same authors conducted a second study~\cite{kanich2011} in which they investigated multiple spamming campaigns and found an even more profitable picture: spammers would earn, on average, \$50 per purchase and the revenues per spamming campaign would range between \$400,000 and \$1,000,000 per month per affiliate program. In both cases, however, such revenues do not include the costs of additional activities related to traditional spams, such as bullet-proof domain hosting or shipping.

Botnet operators are, as discussed above, key actors in the industry and Stone-Gross et al.~\cite{stonegross2011} estimated their potential profit. They estimated the costs of maintaining the Cutwail botnet infrastructure at \$1,500 to \$15,000 on a recurring basis (with 121,336 unique IP addresses online per day) and concluded that botnet owners must have made between \$1.7 to \$4.2 million in profit in the past approximate year. According to this same study, spam-as-a-service can be purchased at a price of \$100-500 per million of spams or botnets can be rented at a price of \$10,000 per month for 100 million of spam emails.

Rao and Reiley~\cite{rao2012} identified three tactics for estimating spamming revenues sent through botnets: monitor a botnet and a market for spams, hijack a botnet and estimate the number of sales made, or evaluate the number of completed orders on a website related to spam by looking at the sequential order IDs in the URL.

In our work, we follow a new tactic, one that can be added to Rao and Reiley's  \cite{rao2012} typology. The tactic aims at estimating illicit sextortion spam revenues by analyzing associated \bitcoin{} transactions, which are openly accessible on a ledger, also known as the blockchain.

\subsection{Estimating Illicit Bitcoin Revenues}\label{sec:relatedwork_methods}

While Bitcoin's share of the cryptocurrency market is shrinking, it still remains the predominant cryptocurrency encountered in cybercrime investigations~\cite{EUROPOL:2018aa}. Hence, tracking and tracing monetary flows has become a de facto standard forensic method for cybercrime investigations and is now supported by a number of commercial (e.g., Chainalysis\footnote{\url{https://www.chainalysis.com/}}, CipherTrace\footnote{\url{https://ciphertrace.com/}}) and open-source tools (e.g. BlockSci~\cite{Kalodner:2017aa}) and analytics platforms (e.g., GraphSense\footnote{\url{https://graphsense.info}}).

These tools rely on three main complementary techniques: first, they apply various clustering heuristics to group multiple addresses into maximal subsets (clusters) that can be likely assigned to the same real-world actor. Multiple-input clustering~\cite{Nakamoto:2008,Reid:2013a}, for instance, leverages the idea that each \bitcoin{} address has a private key that is required to sign a transaction and spend the money associated with it. Thus, if two addresses are used as inputs in a transaction, the private keys of the two addresses must be used to sign the transaction and, subsequently, be accepted by the peer-to-peer network and added to the public ledger. These two addresses can thus be attributed to the same real-world entity, known as a \textit{cluster}. A previous study argued that the multiple-input clustering heuristics are surprisingly effective due to the reuse of addresses and report that super clusters covering thousands of addresses typically represent gambling sites, exchanges or large organizations~\cite{harrigan2016}. Moreover, Nick (2015) ~\cite{nick2015} tested the heuristics on 37,000 wallets and found that it successfully guessed, on average, 68.9\% of all addresses belonging to a wallet. Next to that heuristics, other studies have proposed to form clusters based on change addresses (c.f.,~\cite{Meiklejohn:2013a}) or other behavioral patterns~\cite{Monaco:2015a}.

The second technique is to create network abstractions from the underlying blockchain in which nodes represent addresses or real-world actors by their set of addresses (clusters) and edges represent transactions carried out between addresses or clusters~\cite{Reid:2013a,haslhofer2016,Spagnuolo:2014b}. This provides the data structure required for tracking and tracing monetary flows within a cryptocurrency ecosystem. Finally, address nodes in those networks are enriched with so-called \emph{attribution tags}, which are the key for de-anonymizing actors within an ecosystem; and therefore, the most valuable asset in cryptocurrency forensics tasks. An attribution tag associates a single address with some real-world actor, such as the name of an exchange. The strength lies in the combination of these techniques: a tag attributed to a single address belonging to a larger cluster can easily de-anonymize hundreds of thousands cryptocurrency addresses~\cite{Filtz:2017a}.

The effectiveness of the multiple-input clustering heuristics can, however, provide misleading results if transactions were involved in Bitcoin mixing, either through centralized mixing services (also called tumblers) or through CoinJoin transactions. Centralized mixing services act as middlemen and amalgamate different users' funds before creating new transactions in order to confuse the trail back to the funds' original source (c.f.~\cite{moser2013}). A CoinJoin is a feature often supported by wallet providers or dedicated platforms, which allows several users to combine their funds as input of a single \bitcoin{} transaction without sharing private keys~\cite{Moser2016}. Therefore, CoinJoins must be taken into account when analyzing cryptocurrency flows and a number of detection heuristics are already available in tools like BlockSci~\cite{Kalodner:2017aa}. Furthermore, previous studies have shown that mixing services are typically not used in the first step, when victim pay ransom amounts to some payment address controlled by the attacker~\cite{paquet2018}. This suggests that mixing services are used at later stages before cybercriminals want to cash out and turn tainted funds into clean, spendable fiat currencies.

Above tools and techniques have so far been applied in two studies: Paquet-Clouston et al.~\cite{paquet2018} investigated modern ransomware strains that encrypt a user's computer files and asks for a ransom, usually in \bitcoin{}, for the user to recover the encrypted files. They collected a set of \bitcoin{} seed addresses related to ransomware and expanded that set using the multiple-input heuristics. This allowed them to trace ransomware monetary transactions related to 35 ransomware families and to estimate that the lower-bound revenue of these 35 ransomware families, from 2013 to mid-2015, was near \$12.7 million. Concurrently, Huang et al.~\cite{huang2018} extracted \bitcoin{} addresses from malware files related to ransomware and followed a similar methodology. They ended up estimating that the market for ransomware, over a two-year period, was worth near 16 million of US dollars. These two studies are among the first to provide reliable and reproducible measurement on the revenues of a specific cybercrime: ransomware.

In this paper, we will build and expand on the methodologies presented in those studies and investigate over four million sextortion emails. This should provide insight into the money flow and the potential profitability of sextortion campaigns. Furthermore, this study represents a first analysis of the conjuncture between the spamming industry and cryptocurrencies, a mingle that will likely continue in the next years.

%% file: sections/dataset.tex

\section{Data and Methods}\label{sec:dataset}

In the following section, we describe the dataset, including over 4.2 million spams, and the different methods used to filter and cluster these spams in different campaigns. We also provide summary statistics on the information extracted from each spam.

\subsection{Raw Email Acquisition}

Potentially relevant emails for this research were collected by setting up a tailored spam filter in Virusfree's email security solution between \StartCollectionPeriod~until \EndCollectionPeriod. It filtered all emails containing the terms \emph{BTC} or \emph{Bitcoin} in singular or plural form AND were classified as Spam by Virusfree's spam filter AND originated from an IP that is known by Virusfree as being a member of some botnet. Thus, none of the emails in our dataset reached the recipient. Personally Identifiable Information (PII) in emails, such as recipients' email addresses, were masked before being handed over to us. In total, the raw email dataset contained \nEmails~emails. These emails cannot be published for privacy reasons.

\subsection{Email Bucketing}

To cluster similar spams together, we had to group emails with textual similarity into buckets and identify a representative template for each bucket. In short, we had to find sets of emails with a relatively large intersection (high textual similarity). We tackled this problem by developing a simple \emph{email bucketing heuristic} that allowed us to efficiently compute sets of syntactically similar emails and retrieve a representative email template for each bucket. Our heuristic is based on two observations we made upon manual inspection of random email samples: first, attackers introduce intentional spelling or grammar mistakes or language variations, such as synonyms, across the entire email in order to avoid spam filters. Second, the end of the email contained the most important part of the message, where the attacker explained to the victims what to do. Based on these two observations, we defined our heuristic as follows:

\begin{definition} Let $E$ be the set of emails in our raw dataset and $B$ the set of baskets representing similar sextortion emails.

\begin{enumerate}
  \item For each $e \in E$ find a basket $b \in B \subset E$ with the representative email having the same $l$ last words. If there is such a basket, add $e$ to basket $b$; otherwise, create a new basket with $e_t$ as representative template email.

  \item Compute the Jaccard similarity\footnote{The Jaccard similarity between two sets $S$ and $T$ is defined as $J_{S, T} = \big|S \cap T\big| / \big|S \cup T\big|$} between the $l$ last words in all pairs of template emails $e_t$ representing a basket and merge two baskets if they have a similarity higher than a threshold $t$.

\end{enumerate}

\end{definition}

In order to maximize Jaccard similarities within baskets, we tuned parameters $t$ and $l$ by running experiments on 10\% of the sample emails with all possible pairs $(t, l)$ where $t \in \{0.2, 0.3, 0.4, 0.5\}$ and $l \in \{20, 30, 40, 50, 60\}$. We found that parameters $l = 50$ and $t = 0.3$ yielded maximum similarity with most buckets achieving values greater than 0.8. We considered this a satisfactory result since spammers change textual contents in order to avoid spam filters.

\begin{figure}
    \includegraphics[width=\columnwidth]{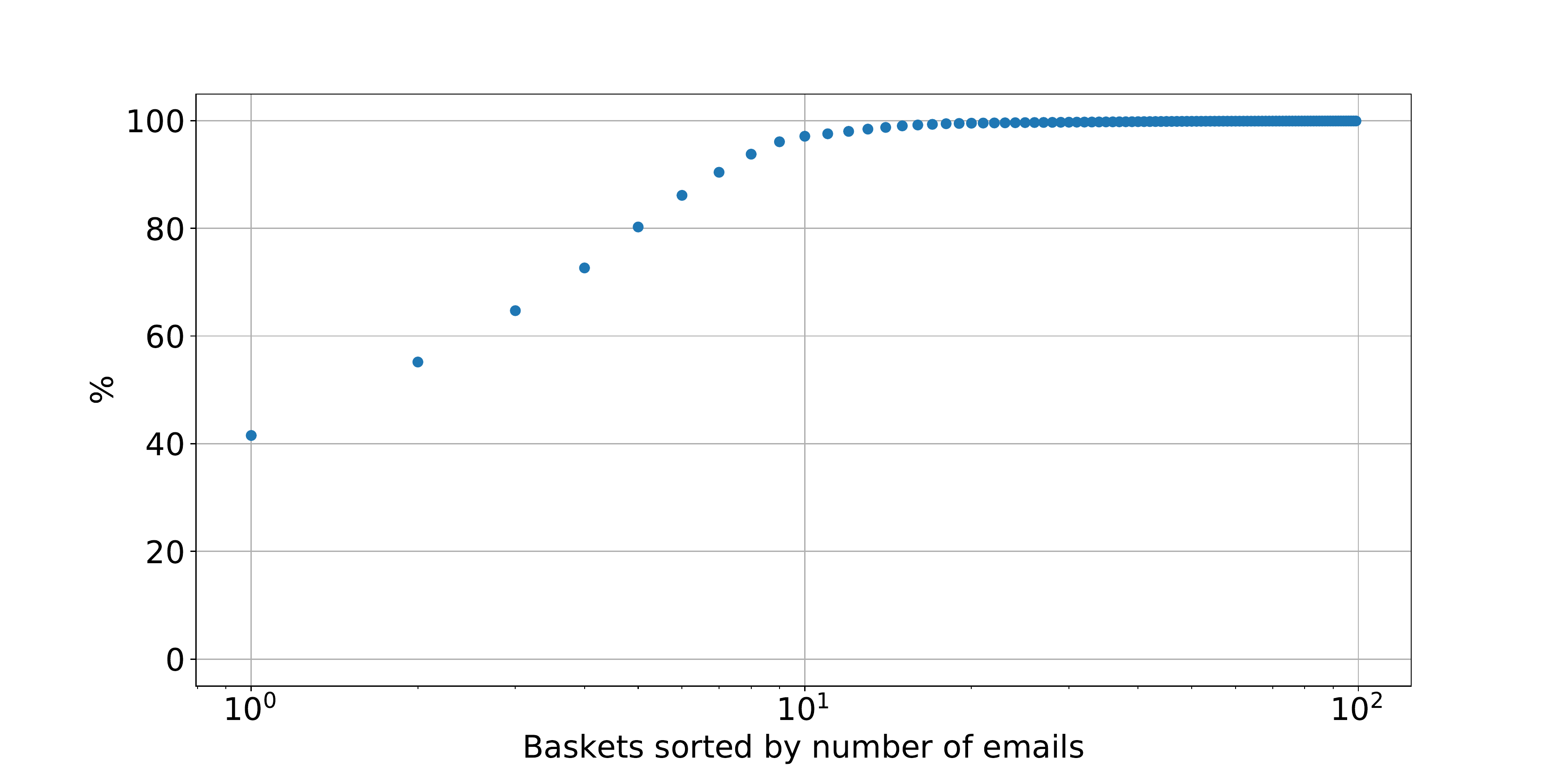}
    \caption{Distribution of emails among the first 100 baskets.}
    \label{fig:emails_baskets_cdf}    
\end{figure}

Overall, our email bucketing heuristic produced \nBasketsStepTwo~buckets with syntactically similar emails, each one having a representative template email. Figure~\ref{fig:emails_baskets_cdf} shows the cumulative distribution of the number of emails in the first 100 baskets ordered by size and shows that 96.1\% of all emails fall into the top 10 buckets.

\subsection{Identifying Buckets Related to Sextortion}

However, at this point, it still remained unclear whether a bucket contained emails related to sextortion or other emails mentioning terms related to Bitcoin. Therefore, we investigated each template representing a bucket and annotated it as being related to \emph{sextortion} or \emph{other}. Since many emails were non-English, we used the Python langdetect\footnote{\url{https://pypi.org/project/langdetect/}} module for automatically detecting the email's language. In case of non-English templates, we used Google Translate\footnote{\url{https://translate.google.com}} to translate their bodies and kept an English translation alongside the original template in our dataset.

This annotation procedure was carried out twice by two independent annotators on all \nBasketsStepTwo~representative email templates. Disagreement on 12 syntactically corrupted email templates was resolved by jointly investigating them in more detail and classifying them as being sextortion. Finally, we ended up with a manually curated corpus of \nEmailsStepFour~ sextortion emails filed into \nBasketsStepFour~buckets representing 99.6\% of the total number of emails shared with us.

\subsection{Categorizing Buckets by Campaign}

Then, we went through each of the \nBasketsStepFour~ templates and found that some of them should be grouped together. Yet, at this point, a manual investigation was necessary and could not be automated. Human interpretation was required because, although the sentence structure and the wording differed among specific templates, the meaning of the message and the nature of the threat were the same and could only be grasped intellectually. This can be illustrated by considering the two following templates:

\vspace{0.25cm}\noindent\textbf{Template No.0 (translated from Czech)}: \textit{Hi! As you may have noticed, I sent you an email from your account. This means I have full access to your account. Your account ([...@....com]) password: [...] I've been looking at you for several months. The fact is that you have been infected with malicious software via the adult website you have visited}.

\vspace{0.25cm}\noindent\textbf{Template No.4 (translated from Czech)}: \textit{Hello! As you may have noticed, I sent you an email from your account. This means that I have full access to your account.  I've been watching you for a few months now. The fact is that you were infected with malware through an adult site that you visited}. 

\vspace{0.25cm}
In template No.0, the attacker has the email and the password of the victim, whereas in template No.4, the attacker doesn't. Yet, the threat that follows is the same with different wording and sentence structure. The buckets representing template No.0 and No.4 should thus be merged, as they encompass the same campaign.

To ensure that our manual investigation was sound, we performed three iterations: going three times manually through each template and grouping them only if specific features around the threat were the same, even though sentence structures and wording differed. Based on this template analysis, a total of \nCampaignsFirst~campaigns, encompassing at least two buckets or more, were found. These \nCampaignsFirst~campaigns include \nBasketsGroupedInCampaigns~of the \nBasketsStepFour~buckets. The 16 remaining buckets that are not grouped still represent a campaign by themselves. Overall, \nCampaignsSecond~distinctive sextortion campaigns were found and they represent our final dataset.

Table~\ref{table:campaigns} presents the specific features of 15 campaigns that were detected in the investigation above. It shows, for example, that Campaign D represents a merged group of templates that mentions the presence of an international hacking group while Campaign G represents another that states that the attacker comes from the DarkNet. Campaign F, on the other hand, details two options to the email recipient: not to pay and live in shame or pay and live safely, while Campaign L mentions the exact \emph{Common Vulnerability Exposure (CVE)} used to infect the device.

\begin{table*}
  \caption{Distinctive features of 15 sextortion spam campaigns}
  \label{table:campaigns}
  \begin{adjustbox}{max width=\textwidth}
    \input{tables/characteristics_campaigns}  \end{adjustbox}
\end{table*}

After having identified buckets related to sextortion and having categorized them by campaign, we extracted the following data points from each email in those buckets: the amount asked\footnote{Amounts were asked in the following currencies: USD, EUR, GBP, NOK} normalized in USD, the time when an email was sent from its \texttt{Date} field in the header, the user's alleged password (or phone number) mentioned in the email, and the Bitcoin address the victim is supposed to send money to. We define all bitcoin addresses found in the spams as \textit{sextortion payment address}. Table~\ref{table:buckets} provides summary statistics on the information extracted from the top 15 baskets representing $99.0\%$ of all emails in our dataset. We also identified four campaigns (not shown in Tables~\ref{table:campaigns} and \ref{table:buckets}) that did not involve hacked email accounts and passwords, but phone numbers.

\begin{table*}
  \caption{Summary statistics on data points extracted from top 15 baskets.}
  \label{table:buckets}
  \begin{adjustbox}{max width=\textwidth}
      \input{tables/short_baskets_table}
  \end{adjustbox}
\end{table*}

\subsection{Finding Sextortion Payment Addresses}

From the \nBasketsStepFour~considered sextortion-related buckets, we extracted \nAddresses~Bitcoin \textit{seed addresses}, but found that only \nExplicitAddressesActive~of them received funds. We consider these \nExplicitAddressesActive~ as being our set of \textit{seed addresses}.

Next, we used the \emph{GraphSense Cryptocurrency Analytics Platform}\footnote{\url{https://graphsense.info}} to expand these seed addresses to maximal address subsets (clusters) computed over the entire Bitcoin blockchain from its inception until April 30th, 2019 (block 573,989). GraphSense applies the well-known~\cite{reid2011anonymity,ron2013quantitative} multiple-input clustering heuristics for that purpose. The underlying intuition is that if two addresses (i.e., $A$ and $B$) are used as inputs in the same transaction while one of these addresses along with
another address (i.e., $B$ and $C$) are used as inputs in another transaction,
then the three addresses ($A$, $B$ and $C$) must somehow be controlled by the
same entity~\cite{Meiklejohn:2013a}, who conducted both transactions and
therefore possesses the private keys corresponding to all three addresses. This heuristic can fail when CoinJoin transactions~\cite{Moser2016} are taken into account, because they combine payments from different spenders that do not necessarily represent
one single entity. Being aware of this problem, we filtered these transactions out using detection heuristics similar to those found in the tool BlockSci \cite{Kalodner:2017aa} before applying the multiple-input heuristics. Also, as discussed in Section~\ref{sec:relatedwork_methods}, we assume that victims do not directly pay into other centralized mixers.

We found that \nAddressesNotInClusters~seed addresses could not be associated with clusters, while the remaining \nAddressesInClusters~appear in \nClusters~different clusters. Table~\ref{table:active_clusters2} reports the number of addresses as well as the first transaction date of the top 10 clusters sorted by amounts received in US dollars. We noticed that two clusters, cluster 0 and 1, differed significantly from the others in terms of the number of addresses, the total amount received and the date of the first transaction. Based on attribution tags retrieved from \emph{walletexplorer.com}, we were able to identify one of these clusters (0) as being related to the exchange \textit{Luno.com}. The real-world identity of the other cluster is unknown, but from its characteristics, we concluded that it represents, most likely, an exchange or another large real-world entity. From these observations, we decided to disregard the addresses included in these two clusters, keeping only their seed addresses.

\begin{table}
  \caption{Summary statistics of top 10 clusters by amounts received.}
  \label{table:active_clusters2}
  \begin{adjustbox}{max width=\columnwidth}
        \input{tables/active_clusters2.tex}
  \end{adjustbox}
\end{table}

Overall, when disregarding the two large address clusters, we expanded our dataset to \nAddressesActive~ address. We assume that the addresses included in this 
\textit{expanded dataset} represent sextortion payment addresses as well. 

\subsection{Filtering Sextortion Transactions}\label{subsec:method_tx}

With \nAddressesActive~sextortion payment addresses at our disposal, we tried to identify Bitcoin transactions representing payment from email recipients to spammers. Since sextortion payment addresses could also be used for other purposes, or other kinds of spamming schemes as discussed in~\cite{schultz2018}, we approached this problem  heuristically and subsequently employed three filtering methods with varying degrees of restrictiveness on all incoming transactions. This allowed us to gradually restrict the set of possible payments to a very narrow, lower-bound estimate.

Inspired by previous work that estimated the ransomware market, while considering potential collectors (addresses used to aggregate money) in the dataset~\cite{paquet2018}, we developed a first filter, dubbed \textit{CollectorFilter}. This filter removes transactions with at least one sextortion address among the inputs and one in the outputs, which indicates money flowing among the group of spammers, and not a victim's payment.

The second filtering method, named \textit{RangeFilter} selects only the transactions where sextortion addresses received payments within the range found in spams, or more precisely, of $ [ (1-p) * \min S , (1+p) * \max S ]$, where $ p = 0.1 $ is a tolerance factor taking into account possible exchange rates oscillations between the email time and the payment time, and $S$ is the set of all the ransom amounts we found in the emails. 

The third filter, dubbed \textit{Moving-MoneyFilter} removes transactions with only one output, also known as \textit{exact-amount payment}~\cite{PrivacyWiki}. Such filter assumes that these transactions represent a sender using the \textit{send maximum amount} feature of a wallet to transfer funds from one address to other, both addresses belonging to the mentioned sender. 



%% file: tables/characteristics_campaigns.tex
\begin{tabular}{cl}
\toprule
Campaign & Key Distinctive Features for each Campaign\\
\midrule
A & A Trojan virus exploited the device and an anti-virus can't detect it as a "Trojan Driver" updates the signature every 4 hours\\
B & RDP was exploited with a keylogger and the device was infected while watching compromising videos\\
C & The primary goal was to lock the device and ask for money, but a sextortion ransom was asked instead\\
D & The attacker is a member of an international hacking group and knows all of the recipient's secrets\\
E & A RAT software was installed on the device and discovered compromising videos\\
F & A virus infected the device via porn websites and two options are presented: pay or not pay \\
G & The attacker is known on the DarkNet and hacked the recipient's mailbox a long time ago \\
H & The attacker is from China and the email is a last warning \\
I & The recipient's password was intercepted via a porn website and a malicious software now intercepts the password every time it is changed\\
J & A Trojan with remote access was installed on the device and the device was additionally hacked through porn websites \\ 
K & The nationality of the attacker is mentioned as an excuse for the poor language structure of the spam \\
L & The attacker claims to be a spyware software developer and mentions the CVE used to compromise the device \\
M & The email recipient's cellphone was compromised and all pictures and videos backed up by the attacker\\ 
N & The email is presented as a serious last warning\\
O & A Trojan infected the device and downloaded all of the recipient's private information, the camera also captured "soloing"\\ 

\bottomrule
\end{tabular}

%% file: tables/short_baskets_table.tex
\begin{tabular}{rllrlllrr}
\toprule
Basket & Campaign & Language &    \#Emails &    Range [\$] & Currency &            Time Period & \#Passwords (Unique) & Jaccard Mean (Var) \\
\midrule
     0 &        A &       cs &  1,811,062 &    280 | 290 &      EUR &  2018-12-02|2018-12-23 &   387,943 (283,828) &         0.98 (0.0) \\
     1 &        E &       cs &    594,904 &    250 | 550 &      USD &  2018-09-06|2018-11-12 &   544,448 (388,809) &         0.96 (0.0) \\
     2 &        B &       en &    415,925 &  300 | 7,000 &      USD &  2018-09-22|2019-01-24 &   414,785 (297,514) &         0.91 (0.0) \\
     3 &        C &       en &    345,969 &    695 | 899 &      USD &  2018-10-29|2019-02-12 &   303,910 (234,354) &         0.92 (0.0) \\
     4 &        A &       en &    331,742 &  500 | 1,000 &  USD,EUR &  2018-11-30|2019-02-13 &               0 (0) &        0.82 (0.01) \\
     5 &        G &       cs &    256,252 &    200 | 290 &      USD &  2018-09-27|2018-11-06 &   256,252 (189,173) &         0.98 (0.0) \\
     6 &        D &       cs &    186,723 &    250 | 250 &      USD &  2018-09-03|2018-11-24 &               0 (0) &          1.0 (0.0) \\
     7 &        F &       en &    146,709 &    700 | 750 &      USD &  2018-12-20|2019-02-15 &    111,114 (97,485) &         0.94 (0.0) \\
     8 &        L &       en &    100,395 &    700 | 999 &      USD &  2018-10-14|2019-01-15 &     54,883 (49,721) &        0.86 (0.01) \\
     9 &        C &       en &     44,790 &  500 | 1,000 &      USD &  2018-09-09|2018-12-27 &     44,454 (26,920) &         0.93 (0.0) \\
    10 &        A &       de &     19,850 &    336 | 784 &      EUR &  2018-12-07|2019-01-21 &               0 (0) &         0.99 (0.0) \\
    11 &        E &       ko &     19,683 &    400 | 500 &      USD &  2018-09-14|2018-12-26 &               0 (0) &          1.0 (0.0) \\
    12 &        G &       en &     18,314 &  200 | 2,000 &      USD &  2018-09-18|2019-02-13 &               0 (0) &        0.79 (0.03) \\
    13 &        C &       de &     13,732 &    336 | 504 &      EUR &  2018-11-24|2019-01-22 &               0 (0) &         0.99 (0.0) \\
    14 &        F &       en &     12,476 &  500 | 7,000 &      USD &  2018-09-13|2019-01-15 &      12,213 (1,472) &         0.45 (0.0) \\
\bottomrule
\end{tabular}

%% file: tables/active_clusters2.tex
\begin{tabular}{rrrrr}
\toprule
Cluster & \#Seed Addresses & \#Addresses & \ Amount Received [\$] & First Tx \\
\midrule
0 & 1 & 398,777 & 2,155,478,411 & 2013-03-22 \\
1 & 3 & 564,595 & 2,110,090,570 & 2016-03-21\\
2 & 16 & 52 & 240,221 & 2018-09-25 \\
3 & 15 & 25 & 200,146 & 2019-01-07 \\
4 & 13 & 22 & 178,245 & 2018-10-30 \\
5 & 6 & 19 & 108,652 & 2018-10-25 \\
6 & 13 & 37 &  103,660 & 2018-09-12 \\
7 & 8 & 29 & 77,062 & 2018-09-10 \\
8 & 3 & 11 & 50,106 & 2019-01-24 \\
9 & 11 & 47 & 36,661 & 2018-10-15\\
\bottomrule
\end{tabular}

%% file: sections/analysis.tex

\section{Analysis}\label{sec:Results}

After having collected Bitcoin addresses and transactions related to sextortion campaigns, we start our analysis by analyzing spammer's pricing strategies. Then we examine reuse of Bitcoin addresses across campaigns, analyze the accumulation of revenues longitudinally, and provide some deeper insights into the monetary flows involved in sextortion spam.

\input{sections/analysis_pricing_strategies}

\input{sections/analysis_Passwords_and_Public_Data_Breaches}

\input{sections/analysis_address_reuse}

\input{sections/analysis_evolution}

\input{sections/analysis_money_flows}



%% file: sections/analysis_pricing_strategies.tex

\subsection{Spammers' Pricing Strategies}\label{subsec:pricing_strategies}

Sextortion spams in our dataset have various threat structures, known as campaigns, and are written in different languages. We hypothesized that sextortion spammers could follow different pricing strategies, based on the language in which the spam is sent and/or the content of the campaigns, as well as whether the email contains personal information, such as the recipient's password or phone number.

First, to systematically assess whether spammers consider language in their pricing strategy, we tested the following \emph{null hypothesis} between language pairs: \emph{the mean amount asked in a certain language does not statistically differ from other languages}. We computed pairwise t-tests, comparing ransom amounts asked based on the language in which the spam was sent. For each language, we conducted normality tests on amounts asked as discussed in~\cite{lumley2002importance}. If, on average, the random samples tested followed a normally shaped distribution, we kept the samples to compute pairwise t-tests. If not, we concluded that the samples were too small in size for the t-tests to be robust to non-normality. We thus disregarded them in the analysis below to avoid doing a Type 1 error (wrongfully rejecting a null hypothesis) when comparing means of amounts asked. Moreover, since variances among spams grouped per language and per campaign showed heteroscedasticity, we applied Satterthwaite correction on sample variances by computing the Welch-Satterthwaite t-test, a test that is designed for unequal variances~\cite{zimmerman1993rank,overall1995tests}. Lastly, to adjust the p-values for multiple comparisons we used the conservative Bonferroni correction.

\begin{figure*}
  \centering
  \subcaptionbox%
    {Overall\label{fig:mean_payment_amounts_overall}}%
    {\includegraphics[width=0.33\textwidth]{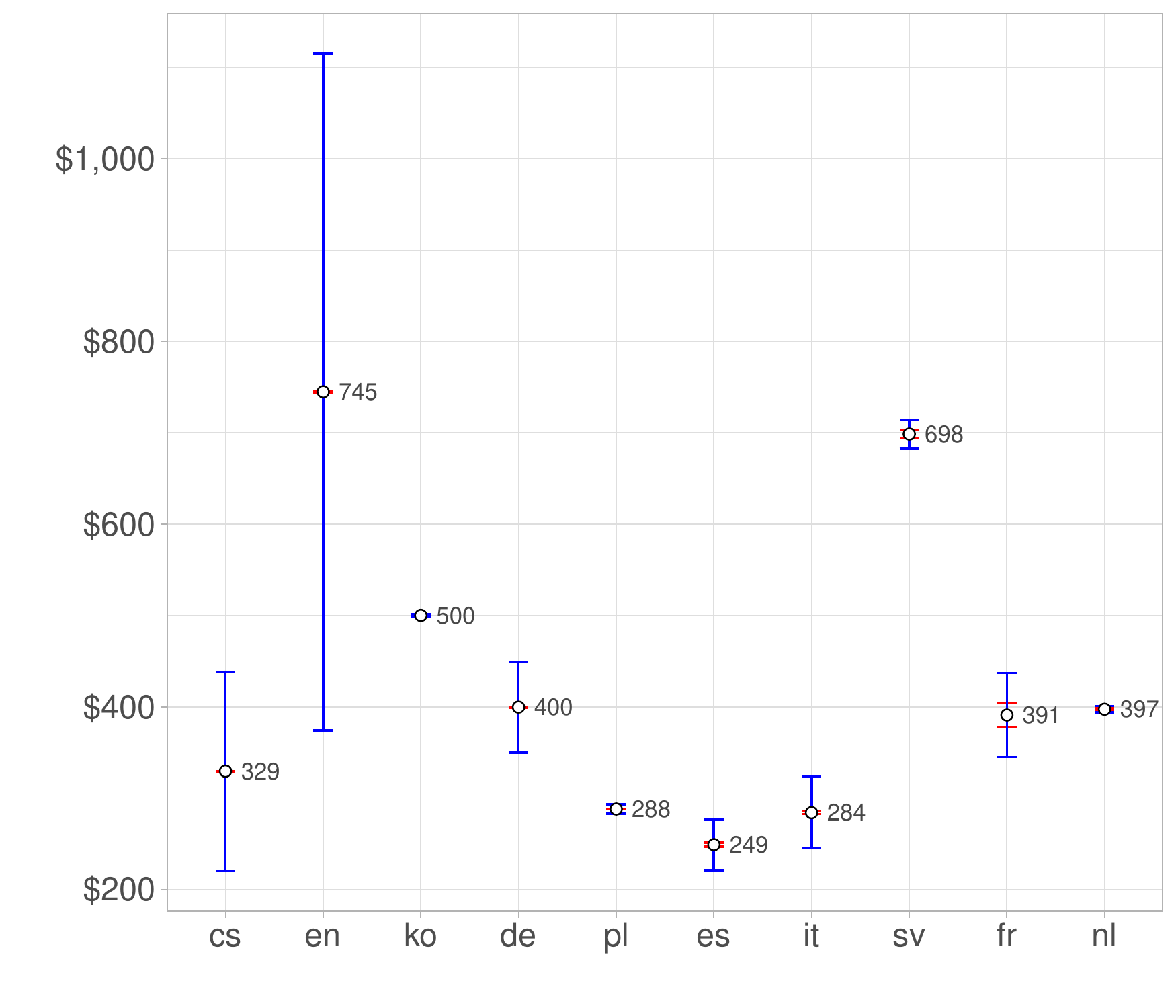}}%
  \hfill
  \subcaptionbox%
    {Campaign A only\label{fig:mean_payment_amounts_CampaignA}}%
    {\includegraphics[width=0.33\textwidth]{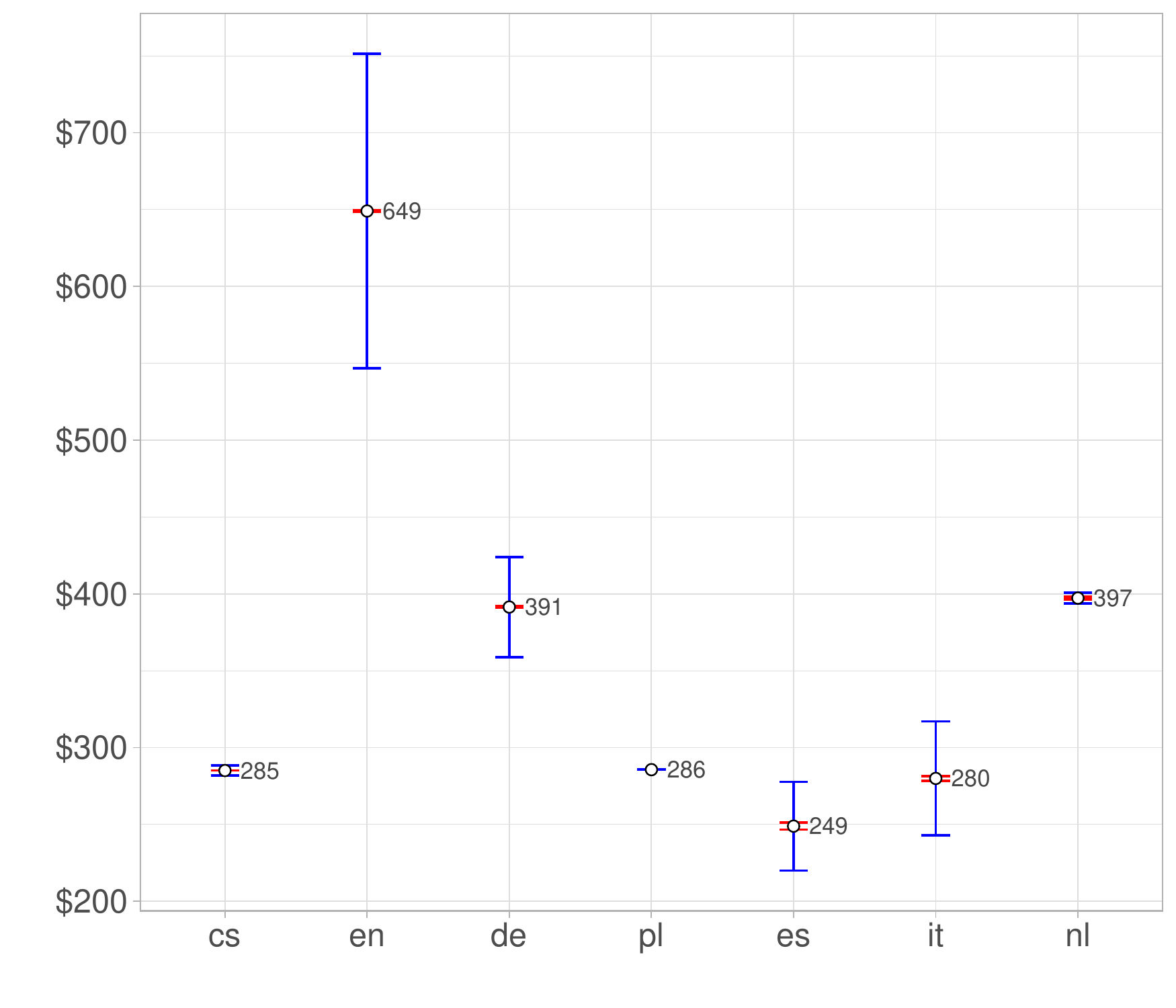}}%
  \hfill
  \subcaptionbox%
    {English only\label{fig:mean_payment_amounts_English}}%
    {\includegraphics[width=0.33\textwidth]{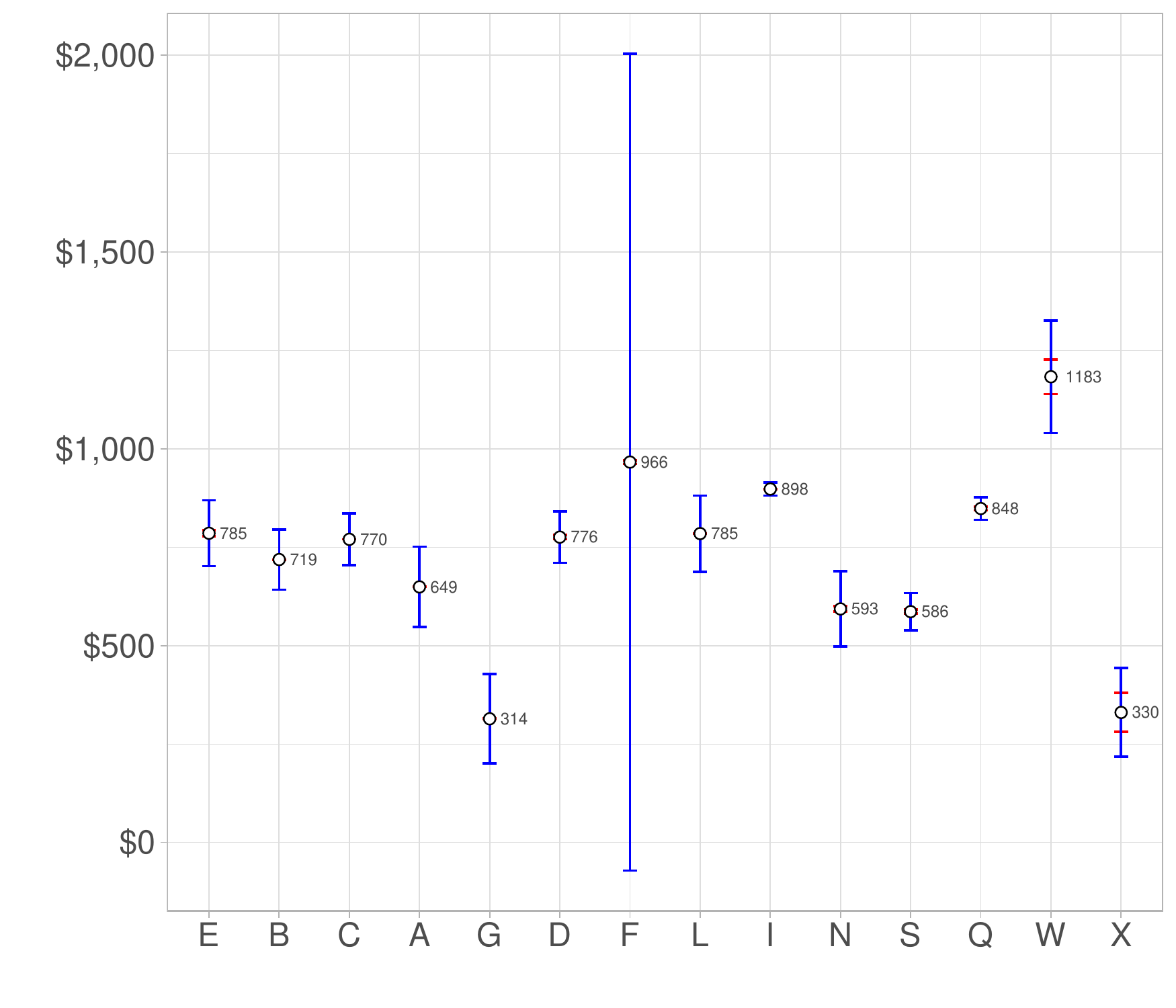}}%
  \caption{Mean amounts asked per Language and per Campaign}
  \label{fig:mean_payment_amounts}
\end{figure*}

The pairwise t-tests revealed that the means of amounts asked per language are significantly different for most of the pairwise comparisons. To discuss these results, Figure~\ref{fig:mean_payment_amounts} presents the mean amounts asked per language, along with the standard deviations in blue and the 1.96 standard mean errors in red. Overlaps of standard mean error bars indicate that the difference between means is zero. 

If we look at pairwise comparisons over all campaigns and languages in Figure~\ref{fig:mean_payment_amounts_overall}, we find that the means of amounts asked in English spams are significantly higher than the means of amounts in all other languages, except Slovenian. We can also observe that the means of amounts asked in German, French, and Dutch are not significantly different. In the lower range of mean payments asked, we find Polish, Italian, and Spanish, which do not significantly differ within their group, but do significantly differ from other languages demanding higher amounts. Japanese, Slovakian and Norwegian spam samples did not pass the normality tests discussed above. Overall, spammers ask higher amounts, on average, for spams sent in English, Slovenian and Korean and lower amounts, on average, for spams sent in Polish, Italian and Spanish.


Next, in order to assess if this finding was robust within one single campaign, we looked at amounts asked per language in Campaign A, a campaign that contains six languages and more than two million spams. Based on Figure~\ref{fig:mean_payment_amounts_CampaignA} and pairwise t-tests between each language in Campaign A, we find that the means of amounts asked in English are significantly higher than all other languages. The means of amounts asked in Dutch are also significantly higher than all other remaining languages. The means of amounts asked in Czech spams are significantly higher than Spanish and Italian spams, but are not significantly different with Polish spams. Overall, we find similar results, with recipients of English spams facing higher amounts asked than recipients of spams in all other languages. Also within Campaign A, Spanish spams asked the cheapest amounts, on average. 

We looked if amounts asked differed per campaign. To avoid grasping a language effect, we selected only English spams, a language encompassing 1,427,311 spams and 31 campaigns. Then, we selected the 14 campaigns that passed the normality test mentioned above. As shown in Figure~\ref{fig:mean_payment_amounts_English} and based on pairwise t-tests, we find that there are significant differences in the means of amounts asked depending on the campaign. For example, the means of amounts asked in Campaign G are significantly lower than all other campaigns except Campaign X. Amount asked in Campaign W are also significantly higher than amounts asked in all other campaigns. Campaign F also shows that amounts asked in English vary greatly. Overall, we find that there are significant differences in the amounts asked per campaign.

Lastly, we found no significant differences between spams sent with and without the recipient's password or phone number, even when controlling for the language in which the spam was sent and the campaign in which it was classified. Spammers do not ask for higher amounts when they have additional information from the recipient. In summary, from these findings, we can conclude that sextortion spammers consider language in their pricing strategy within and across campaigns. They also ask for different amounts depending on the campaigns.

%% file: sections/analysis_Passwords_and_Public_Data_Breaches.tex

\subsection{Passwords and Public Data Breaches}

Schultz~\cite{schultz2018} hypothesized that the emails and passwords used by sextortion spammers come from publicly available data breaches. To explore this hypothesis, we selected all unique clear text passwords (as email addresses were masked for privacy purposes) from the spamming dataset. We then created a list of clear text passwords that were seen only once in the dataset, to avoid passwords such as \textit{12345}, and that comprised at least four characters, to avoid passwords such as \textit{!!}.

This yielded a list of 632,688 unique passwords. We fetched 21 publicly available data breaches (e.g., MySpace, Linkedin, Twitter, Yahoo, 665KK), encompassing nearly 800 million passwords. Due to computational limits, we randomly selected 25\% of the passwords included in the list above and investigated whether they were contained in one of these breaches. A total of 80,944 passwords were found to be a match. In short, about 51\% of our randomly selected password samples used by sextortion spammers were passwords found in previous clear text data breaches. These findings strongly support the idea that sextortion spammers use publicly available data breaches to populate the content of their spams.

%% file: sections/analysis_address_reuse.tex

\subsection{Reuse of Bitcoin Addresses across Campaigns}\label{subsec:analysis_reuse}

When investigating Bitcoin addresses extracted from the buckets listed in Table~\ref{table:buckets}, we noticed that addresses and associated clusters are reused across buckets and sextortion campaigns, confirming similar observations made by researchers previously ~\cite{schultz2018, ibmx-forceexchange2018}. More specifically, we observed that \nClustersWithMultpleBaskets{} clusters appear in at least two distinct baskets, and from this, we hypothesized that the financial revenue streams of a large majority of sextortion campaigns are controlled by a single real-world entity despite their semantic and language differences.

We evaluated our hypothesis by creating an undirected graph abstraction from the \nBasketsStepFour{} buckets that were identified as being related to sextortion campaigns. Nodes in the graph represent buckets while edges between nodes indicate shared clusters or shared seed addresses among two buckets. Thus, an edge between two nodes indicates that access to funds collected from sextortion victims is controlled by the same real-world entity, holding the private keys of some addresses included in both nodes.

\begin{figure}
    \includegraphics[width=\columnwidth]{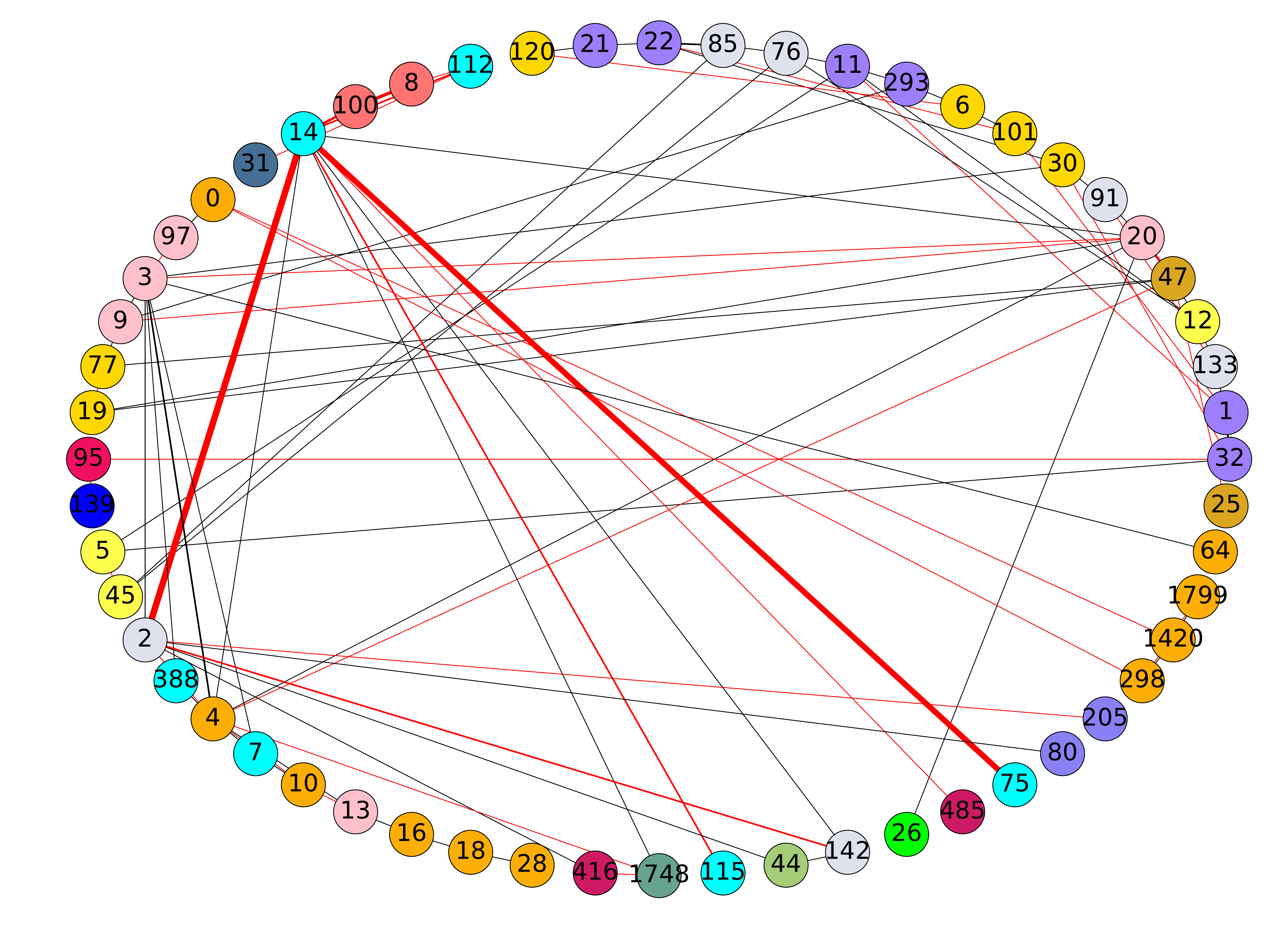}
    \caption{Relationship among sextortion buckets, campaigns, bitcoin addresses and clusters.}
    \label{fig:baskets_connections}    
\end{figure}

When analyzing that graph abstraction, we found that the overall graph consists of 4 disconnected components, with one giant component covering 56 out of \nBasketsStepFour~nodes, which include 99.4\% of the emails and 72\% of the sextortion addresses. Two other disconnected components have only three nodes while the last component has two nodes. All the remaining nodes are disconnected. Figure~\ref{fig:baskets_connections} represents the giant component and shows the financial connections between buckets and campaigns identified by overlapping \bitcoin{} clusters or addresses. Red edges indicate that the connected buckets share at least one \bitcoin{} address. Black edges denote intersections in the set of addresses (clusters) associated with two connected bucket nodes. The strength of an edge indicates the number of shared entities (either clusters or addresses) and node colors represent campaigns.

In summary, our analytics strongly indicates that a large majority of the financial revenue streams generated by sextortion spam can essentially be attributed to a single real-world entity. We note that this does not necessarily imply that all sextortion campaigns are designed and carried out by a single person. It is more likely that botnet infrastructures, which can be rented by spammers, are nowadays designed with financial features that handle the creation of \bitcoin{} addresses, and potentially handle the financial revenue streams within the \bitcoin{} ecosystem. However, more research on the matter is needed. 

%% file: sections/analysis_evolution.tex

\subsection{Sextortion Revenues}\label{subsec:analysis_evolution}

Next, in order to create reliable and reproducible statistics on the profitability of sextortion spam, we estimated the profitability of sextortion spams based on the emails contained in our dataset. Since our observation period covers five months (\StartCollectionPeriod{} until \EndCollectionPeriod{}), we also wanted to understand how the revenue stream evolved over time. 

To gain insight into the overall profitability, we first quantified revenues with various filter settings: we applied the \textit{CollectorFilter} (1) and the \textit{RangeFilter} (2) presented previously on all incoming transactions of the expanded dataset, providing a reasonable revenue estimation. Additionally, we applied the \textit{Moving-MoneyFilter} (3). However, we considered the estimate including the third filter as a \textit{conservative one}. Indeed, the \textit{Moving-MoneyFilter}  could disregard potential transactions related to sextortion, as victims could send money to a sextortion address without providing a change address.

Table~\ref{table:revenues} shows the number of payments, as well as the total amounts received by the sextortion addresses based on the different filtering criteria presented above. It shows that the overall revenue of sextortion spams lies between \$\revenuesThirdFilter{} and \$\revenuesSecondFilter{}.

\begin{table}
  \caption{Number of received payments and estimated revenues using different filtering criteria.}
  \label{table:revenues}
  \input{tables/filters_revenues}
\end{table}

\begin{figure}
    \includegraphics[width=\columnwidth]{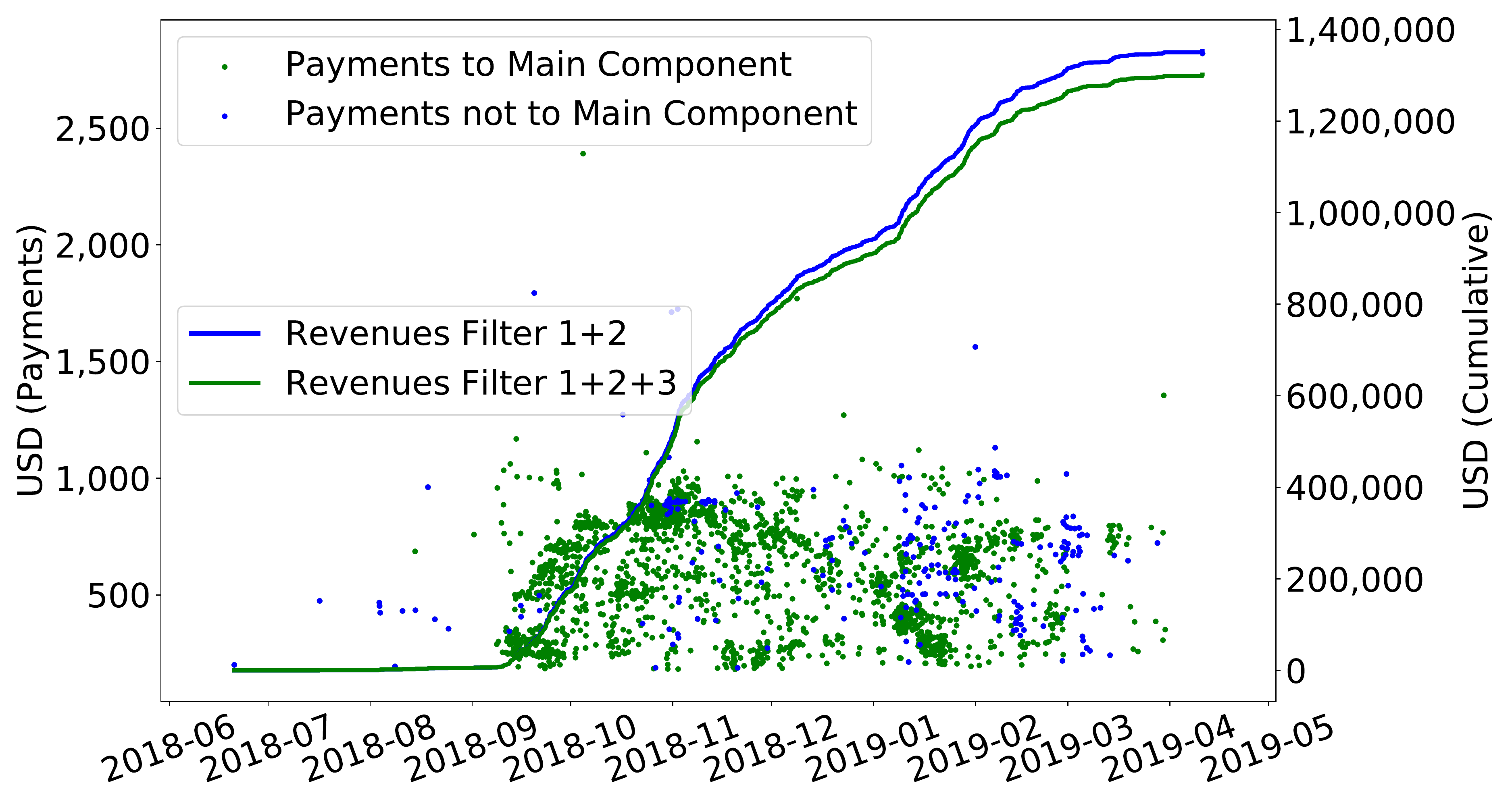}
    \caption{Sextortion's cumulative revenues combining different filters and single payments to sextortion addresses.}
    \label{fig:payments_scatter}
\end{figure}

Figure~\ref{fig:payments_scatter} illustrates all the individual payments to sextortion addresses and the cumulative sum over time. Although we collected spams from September 2018 to March 2019, Figure~\ref{fig:payments_scatter} shows that we captured payments before and after this period. The revenue estimations thus include payments between June 2018 and April 2019.  Figure~\ref{fig:payments_scatter} also shows that sextortion addresses related to the one single component, discussed in the previous section, accounts for 88.3\% of the overall sextortion revenues. Moreover, from September 2018 to March 2019, we find that sextortion addresses receive an almost steady flow of payments, with slightly higher revenues in October, November and January with \$\revenuesOctober{}, \$\revenuesNovember{} and \$\revenuesJanuary{} respectively. The average monthly revenue being \$\averageMonthlyRevenue.   

%% file: tables/filters_revenues.tex
\begin{tabular}{rrrr}
\toprule
Filters & \#Payments & Revenues [\$] & Revenues [BTC] \\
\midrule
    1+2 &     2,346 &    1,352,266 &        281.776 \\
  1+2+3 &     2,252 &    1,300,620 &        269.983 \\
\bottomrule
\end{tabular}

%% file: sections/analysis_money_flows.tex

\subsection{Holding Periods and Monetary Flows}

To better understand when and how victims' coins change hands, we investigated the holding periods of received payments and subsequent monetary flows to other addresses or clusters.

\begin{figure}
    \includegraphics[width=\columnwidth]{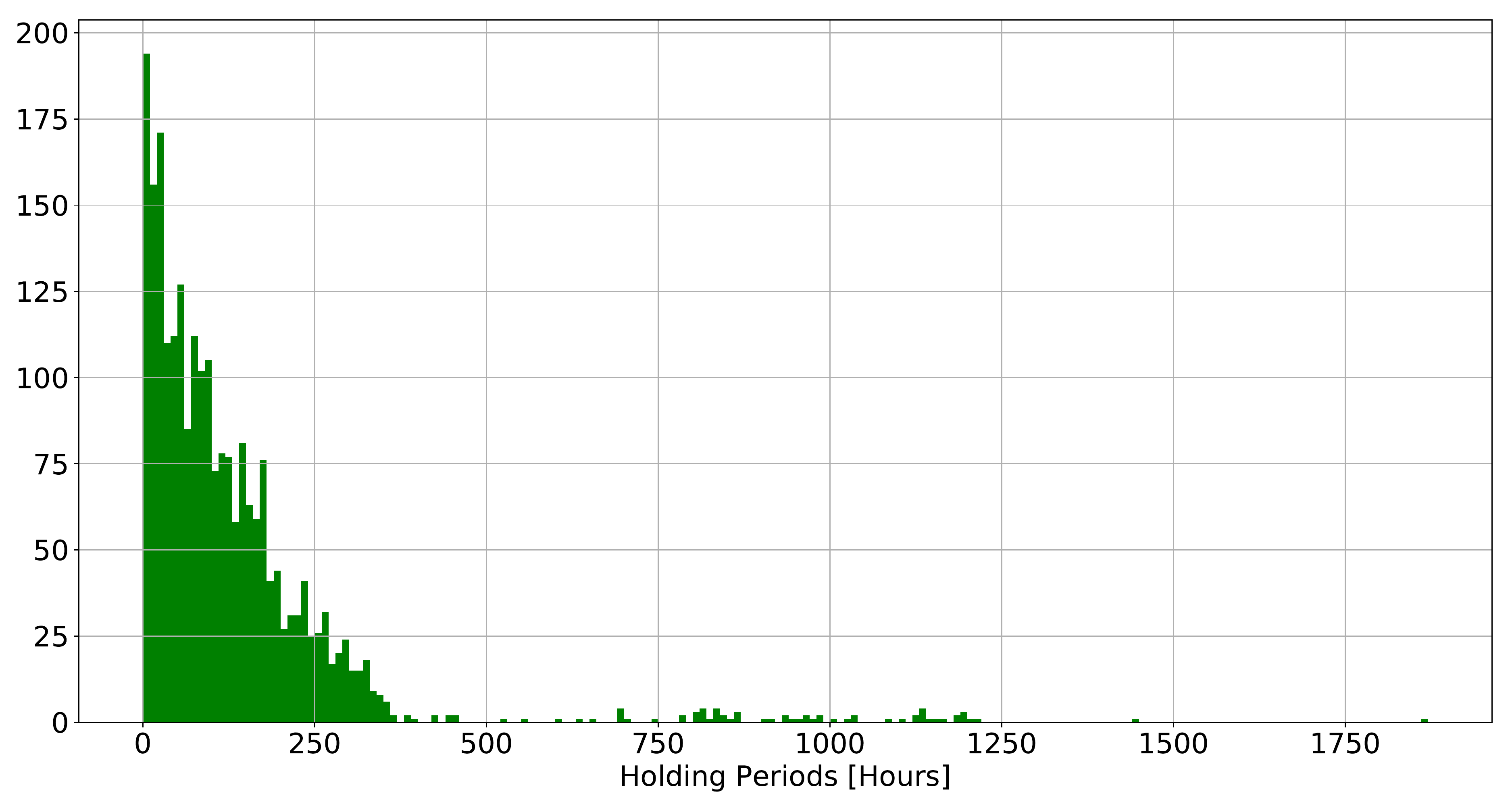}
    \caption{Histogram of holding periods considering the amounts received by sextortion addresses. Each bin has a width of 10 hours.}
    \label{fig:holding_periods}    
\end{figure}

A holding period is the time elapsed between the moment a sextortion address received coins from a victim and the moment these coins are spent. We found that holding periods vary between few minutes and few months, with an average holding time of 5.5 days and a median of 3.8 days, as shown in Figure~\ref{fig:holding_periods}.

Then, we investigated monetary flows that start from a victim's address (depth 0) and forward amounts to other addresses through several hops. Technically, this corresponds to traversing a transaction graph, in which nodes represent addresses and edges represent transactions. In order to limit computational efforts, we stopped traversing when we hit an address representing a known exchange or when we encountered more than 100 nodes at a certain depth.

For the analysis, we only considered transactions that have been flagged as sextortion payments, based on the \textit{Collector-} and the \textit{Range} filters, yielding \nTxsSecondFilter{} sextortion payments related to \nAddressesNotInClustersSextortion{} addresses and \nClustersSextortion{} clusters. We did not consider the \textit{Moving-Money} filter because it might exclude payments where the victim does not care about the change.

We found that all the coins sent from the victims to the sextortion addresses have been moved to other wallets. Within two hops, we already found entities representing known cryptocurrency exchanges, such as Huobi.com, CoinPayments.net, Cubits.com, and again Luno.com, thanks to the attribution tags retrieved from \emph{walletexplorer.com}. These known entities, however, only received 3.8181 BTC, representing about 0.135\% of the total estimated sextortion revenues. Such finding would indicate that spammers do not necessarily cash out within the first hops. However, since our attribution dataset is limited to \textit{walletexplorer.com}, we cannot conclude with certainty this assumption, as untagged clusters may represent exchanges or money-laundering services that we are not aware of.

As a first explanatory exercise, we present in Table~\ref{table:big_clusters_depth2} all unknown clusters found at depth 2, meaning the clusters that received money from known sextortion addresses. As shown in this Table, some clusters spent millions of dollars, were active years before sextortion spams were first seen, and completed thousands of transactions. Considering these features, such clusters could represent exchanges or money-laundering services. As an extra exercise, we computed the proportion of sextortion revenues that were moved to clusters that have a transaction that is older than June 2018. Such criterion -June 2018- represents the month in which we observed a first sextortion payment in our dataset. Clusters being active before June 2018 could represent potential cash outs, since they are older than the spammers' observed activities. We find a total of 48.218340 BTC, 17\% of spammers' revenue going through these clusters, representing potential cash outs. However, this analysis is arbitrary and further research is needed to determine what features of a cluster (e.g. total amount spent, number of addresses, activity period) can hint an individual that the cluster in which the coins are flowing, although untagged, is a potential cash out service. 

\begin{table}
  \caption{Top 15 clusters by BTC received from victims at depth 2.}
  \label{table:big_clusters_depth2}
  \input{tables/big_clusters_depth2.tex}
\end{table}



%% file: tables/big_clusters_depth2.tex
\begin{tabular}{rlllr}
\toprule
   Cluster & Total Spent [\$] &    First Tx &   \#Txs Out &        BTC \\
\midrule
 466805936 &         113,082 &  2019-01-09 &          1 &  19.324660 \\
 374981799 &      52,789,437 &  2017-12-27 &        179 &  18.418278 \\
 363386240 &   2,110,090,569 &  2016-03-21 &    377,671 &  10.377687 \\
 452279937 &          16,690 &  2018-11-19 &          1 &   3.781298 \\
 426830314 &          36,454 &  2018-07-23 &         20 &   3.513247 \\
 477564438 &          17,636 &  2019-01-20 &          1 &   3.476584 \\
 329777716 &     154,186,706 &  2013-04-16 &    226,156 &   2.977402 \\
 484209796 &     488,989,966 &  2018-08-04 &    159,198 &   2.959614 \\
 498338351 &  27,513,490,714 &  2014-09-04 &  3,150,371 &   2.945446 \\
 310058166 &       6,016,682 &  2016-10-28 &        633 &   2.180812 \\
 490354283 &   1,344,877,122 &  2018-10-04 &     13,230 &   1.764936 \\
 469936395 &          11,736 &  2019-01-15 &          1 &   1.654278 \\
 468575380 &          33,950 &  2019-01-09 &         22 &   1.544722 \\
 496671952 &       3,421,775 &  2019-01-11 &        129 &   1.517212 \\
 386226499 &       9,883,607 &  2016-12-07 &        565 &   1.420000 \\
\bottomrule
\end{tabular}

%% file: sections/discussion.tex

\section{Discussion}\label{sec:discussion}

In this section, we first discuss our findings about the limited sophistication of sextortion spammers. We then provide estimates of spammers' revenues and profits and lay down the limits of this research. 

\subsection{The Somewhat Sophistication of Sextortion Spam Operations}

This study finds that sextortion spammers developed a large array of threat strategies to convince spam recipients to pay the amount asked: a total of \nBasketsGroupedInCampaigns~ different campaigns were found. The amount asked also differed based on the campaign in which a spam was included into. However, we couldn't find a precise reason as to why a campaign would display higher average amounts compared to other campaigns. Identified features, such as Campaign W stating that both the device and the spam recipient's phone were exploited, or campaign G, mentioning that the attacker comes from the DarkNet, were insufficient to find why some campaigns would be asking for higher amounts compared to others.  

We also found that sextortion spammers adjust the price of spams to the language in which the spam is sent. Our results showed that, on average, English (\$745), Slovenian (\$698) and Korean (\$500) are charged higher amounts than Polish (\$288), Italian (\$284) and Spanish (\$249) recipients. This is in line with other related results: Hernandez et al.~\cite{hernandez2017} found that ransomware authors charged their victims differently depending on many variables, including language. Another study found malware-as-a-service was sold at a higher price on English websites compared to Russian ones~\cite{gutmann2007}. Thus, our study finds that, similar to other illicit online activities, spammers also conduct price discrimination based on language. However, explanations as to why the language of a spam recipient changes the amount asked by sextortion spammers are yet to be discovered. There are many potential arguments, such as testing different threat strategies or subjectively deciding that a specific language-speaking population should be charged more. However, such analysis is beyond the scope of this paper. 

We also found that spammers do not charge higher prices for spams that include passwords and/or phone numbers. This is surprising considering that these additional elements could make the threat more convincing. Moreover, spammers do not seem to track down which threat strategy is more efficient or which language-speaking population pays more, as \bitcoin{} addresses are being reused among spams, regardless of the campaign or the language in which the spam is sent. Such findings indicate two tendencies: spammers do develop strategies related to the spam content, but they do not track afterwards the success of their operations (which campaign paid more, for example). Their sophistication is thus, limited. 

Potentially, the fact that spammers do not track payments could be related to the idea that botnet infrastructures are nowadays designed with financial features that handle the creation of \bitcoin{} addresses, and possibly handle the financial revenue streams within the \bitcoin{} ecosystem. The sophistication would thus be limited to pre-spamming activities. 

\subsection{Revenues and Costs of Sextortion Spams} 

Our research results indicate that one single entity is likely controlling the financial backbone of the majority of sextortion campaigns investigated in this study. Moreover, depending on the applied filter, sextortion spams generated a lower-bound revenue between \$\revenuesThirdFilter{} and \$\revenuesSecondFilter{} for an 11-month operation, or an average of \$ \averageMonthlyRevenue{} per month, between June 2018 and April 2019 when payments were captured.

Traditional spam was found to be as lucrative, with findings on revenues ranging from 3.5 million per year~\cite{kanich2008} to between \$400,000 to 1 million a month~\cite{kanich2011}, depending on the study. However, as discussed above, sextortion spams cut the upper-tail of the supply chain compared to the traditional spamming industry~\cite{levchenko2011}. The costs of sending sextortion spams are much lower than that the costs of sending traditional spams related affiliate marketing~\cite{rao2012} or the sale of pharmaceutical products~\cite{kanich2008,kanich2011} or counterfeit goods~\cite{stringhini2014}, as no third-party relationships are needed, nor product shipments or hosting of illicit websites, for example. 

Moreover, a large proportion of the passwords found in the dataset were related to available data leaks and pastes. Sextortion spammers are thus likely to use emails found online, through public data breaches, rather than buy email lists from harvesters, as argued in previous research~\cite{john2009,kanich2011,stonegross2011}. The price of harvesting these emails would thus be quite limited.    

At another level, all evidences point to the idea that sextortion spammers are using botnets to disseminate their spams, just like traditional spammers. Indeed, two previous articles on sextortion~\cite{schultz2018,ibmx-forceexchange2018} argued that this worldwide campaign was sent by the Necurs botnet. This assumption was based on the geolocations of the IP addresses sending the spams, which were similar to the Necurs's infected hosts. Our sextortion dataset was also created partly through a filter that selected emails sent from IP addresses that were known to be part of botnets, supporting this hypothesis as well. 

Moreover, according to Stone-Gross et al.~\ \cite{stonegross2011}, spam-as-a-service can be purchased at a price of \$100-500 per million of spams or \$10,000 per month for 100 million of spams. Kanich et al.~\cite{kanich2008} also found that 350 million of spams soliciting pharmaceutical products led to 28 sales. This means that 12.5 million of spams would need to be sent for one sale to be conducted. Hypothesizing that the lower-tail of the spamming industry is segmented \cite{stringhini2014} and that sextortion spammers operated within these eleven months by renting a botnet at a price of \$10,000 per month for 100 million of spams~\cite{kanich2008}, a practice that was found in several previous research \cite{john2009, stonegross2011, kanich2011}, sextortion spammers would have paid a total of \$110,000 to botnet owners. This means that they would have made a lower-bound profit of \$1,190,620 and \$ 1,242,266 for eleven months. Such estimate, however, does not include, other costs, such as the costs of crafting the spams or of cashing out through the \bitcoin{} ecosystem. Yet, this potential profit magnitude shows that sextortion spamming is a lucrative business that is likely to continue, especially considering that steady flows of payments every month were observed in our dataset.

\subsection{Limits and Future Research}

This research faces some limits that can drive forthcoming studies. First, the dataset studied is based on emails that never reach the intended recipients. The lower-bound estimates are derived from the reuse of \bitcoin{} addresses by spammers in their sextortion spams. Potentially, the reported numbers underestimate the magnitude of the spammers' revenues. Thus, possibly, our dataset represents only a small proportion of a global campaign. The multiple-input heuristic allowed us to find many addresses related to sextortion, but we cannot confirm that the expanded dataset is holistic. Future research could reproduce our methodology with additional addresses. Upcoming studies could also look into the potential conversion rate of sextortion spams, a measure that we could not find due to our limited dataset. 

Second, the address clustering method we applied in our study is a best effort approach based on behavioral patterns such as reuse of addresses across transactions. We cannot preclude that it unifies entities that have no association in the real world, or that it misses to unify entities. Our method also relies on the availability and reliability of attribution tags, which are currently harvested from public sources (mainly \emph{walletexplorer.com}) without a verification process in-between. Future investigations that aim at understanding the real-world actors involved in sextortion campaigns and associated monetary flows clearly have to provide safeguards for ensuring the evidential value of such kind of analysis, by recording, for example, the provenance of attribution tags. Moreover, the tracking of money flows is limited by these attribution tags. Upcoming research should look into determining clusters' features that can detect a potential cash-out service, without the use of the attribution tag, allowing the tracking to stop once such a cluster is found. 

The expanded dataset may also include transactions that are not related to sextortion spams. Indeed, Schultz~\cite{schultz2018} did find one example of an address being reused in sextortion spam and in a scam where a Russian girl would send a video in exchange of a small \bitcoin{} amount. Such spams were not found in our dataset nor was the example provided by Schultz (2018) found in the campaigns studied. Yet, we cannot exclude the possibility that the revenues estimated include other kinds of spamming schemes that aim at influencing spam recipients to send money to addresses. Subsequent studies should aim at understanding other schemes developed by spammers to take advantage of cryptocurrencies and extract money from spams recipients, an amalgamation that is unlikely to go away.


%% file: sections/conclusions.tex

\section{Conclusion}\label{sec:conclusions}

This study shows that spammers have upped their game: mixing the cryptocurrency bitcoin with spams by trying to extort money from spam recipients. This is done through a false threat: compromising videos or photos will be sent to the recipient's contacts if an amount of money is not sent to a specific bitcoin addresses. We find that sextortion spammers are somewhat sophisticated, following specific pricing strategies, based on language and threat structures. Yet, our findings also indicate that they do not track whether their strategies are successful, hence the idea of "somewhat". Moreover, we conclude that sextortion spammers cut the upper-tail of the spamming supply chain, reducing the spamming cost. Tracking their financial transactions in the \bitcoin{} ecosystem, we find that these spammers made a lower-bound revenue between \$\revenuesThirdFilter{} and \$\revenuesSecondFilter{} for an 11-month operation. 

We conclude that sextortion spamming is a lucrative business and spammers will continue to send bulk emails that try to extort money through cryptocurrencies. Other schemes, different than sextortion, but involving cryptocurrencies, will likely be developed by spammers, allowing such industry to thrive. The anti-spam community should thus focus on preventing these spams to reach users' inbox by flagging messages that include cryptocurrency addresses and that are coming from IP addresses that are known to be part of botnets.